# Becoming Agents of Change through Participation in a Teacher-Driven Professional Research Community

Mike Ross, Ben Van Dusen, Valerie Otero, University of Colorado Boulder, 249 UCB, Boulder, CO 80309
Email: michael.j.ross@colorado.edu, benjamin.vandusen@colorado.edu, valerie.otero@colorado.edu

**Abstract:** This study involves a theory-based teacher professional development model that was created to address two problems. First, dominant modes of science teacher professional development have been inadequate in helping teachers create learning environments that engage students in the practices of science, as called for most recently by the NGSS. Second, there is a lack of teacher presence and voice in the national dialogue on education reform and assessment. In this study, teachers led and participated in a professional community focusing on STEM education research. In this community, teachers became increasingly responsible for designing and enacting learning experiences for themselves and their colleagues. We investigated the characteristics of the science teachers' learning process. Findings suggest that teachers who participated in this model generated knowledge and practices about teaching and learning while simultaneously developing identities and practices as education reform advocates and agents of educational change.

## Introduction

The goal of the NSF-funded, Streamline to Mastery project was to develop and investigate a model of science teacher professional development (PD) that is, by design, customized to participating teachers' needs and experiences. Customization is made possible by organizing activity so that teachers increasingly take control of, and responsibility for, their own professional development. Through the task of creating learning experiences for themselves, such as engaging in lesson-sharing and developing their own education research agendas, the nine participating Streamline teachers generated knowledge about teaching and learning. This knowledge took the form of principles of science teaching and learning and propagated in various forms, locally and nationally, to other researchers and practitioners. Workshops for teachers, conference presentations, peer-reviewed publications, and leadership roles in local schools and on national committees have become common means for the Streamline teachers to participate in productive dialogue on science education reform.

Streamline to Mastery was created in response to two problems. The first is the widespread perception that dominant, top-down modes of science teacher PD are largely ineffective (e.g., Borko, 2004). Two of the three authors are former high school physics teachers and come from communities of practitioners that regard top-down PD as costly, disconnected from the needs of teachers, and largely ineffective at instigating meaningful change. Though researchers have theorized about innovative PD experiences (e.g., Grossman, Wineburg, & Woolworth, 2001) and examined factors that appear to make some teacher PD more effective than others (e.g., Borko, 2004; Garet, Porter, Desimone, Birman, & Yoon, 2001), this study seeks to explore a model of teacher learning that deviates significantly from conventional thinking about how changes in practices and broader agency can be realized by science teachers. The second problem that Streamline was designed to address is the conspicuous absence of teachers from the national dialogue on education reform. This dialogue is heavily influenced by groups with little or no connections to actual K-12 classrooms, including policy think tanks, corporate interests, and philanthropic organizations. University researchers make up a large and vocal group of education reformers as well, many of whom are former teachers who conduct research in or about K-12 schools. We argue that even the most well-intentioned of the university researchers are not grounded in the day-to-day reality of being a teacher in the current climate. It is difficult to imagine a conference on best surgical practices being virtually devoid of practicing surgeons, yet that is the unfortunate and unacceptable state of many prestigious conferences in K-12 education research.

In this piece, we present data showing shifts in the roles and practices of the Streamline participants from those typical of practitioners to those associated with teacher leaders and agents of broad educational change. We bring a particular focus to the development and abstraction of general principles of teaching and learning by the Streamline teachers and discuss these practices as they relate to scientific practices and science learning. Finally, we discuss the implications of these findings for the development of future science teacher PD efforts, particularly with regard to the Next Generation Science Standards (NRC, 2013).

## Theoretical Framework

Holland, et al. (Holland, D., Lachicotte, W., Skinner, D., Cain, 1998) define identity as a relational phenomenon as much bound to others and all manner of cultural forms as to the person to whom it is attributed. Sociocultural conceptions of identity have been shaped by the work of Vygotsky (1986), which brought the broader social activity into accounts of individual learning. Contemporary researchers have explored relationships between the constructs of identity, agency, and goals (e.g., Barton & Tan, 2010; Nasir & Hand, 2006; Nasir, 2002) as a means to understand how perceived opportunities and constraints are a medium through which identity and culture mutually and continuously construct one another.

Our work is grounded in the view that identity and cultural practices are mutually constituted. As it relates, we view cultural practices as the aggregate of ever evolving expressions of shared expectations, norms, and values—as expressed through agreed-upon practices—among individuals that are identified by, and identify with, these practices. In this view, each individual navigates and negotiates her own identity development in the world of what is perceived to be possible (or not); meanwhile, this continual negotiation shapes the broader context and thus what is perceived to be possible. Changes in an individual's practices, and in how she positions herself in and through those practices, *constitute* learning.

In this study, we investigate the hypotheses that social contexts in which participation requires authoring of activity (Bahktin, 1993) lead to changes in participation and agency within that social context. For science students, these forms of agency involve capacity and empowerment to engage in scientific practices (Belleau & Otero, 2012; Ross & Otero, 2011; Van Dusen & Otero, 2012). For K-12 science teachers, forms of agency involve capacity and empowerment to instigate change in local science education contexts, such as the teachers' classrooms, as well as the instigation of broader change (Ross & Otero, 2011; Ross, Van Dusen, Sherman, and Otero, 2012). In both settings, we have investigated the hypothesis that learning contexts can be designed to enable learners to become authors of disciplinary knowledge on the basis of collaborative reflection on their experiences. We argue that this agency is associated with identities as scientists (with K-12 students) and instigators of change (with science teachers).

We propose a heuristic of two extremes of K-12 teacher professional development to frame our research: (1) a model in which principles of teaching and learning are provided by experts who administer the PD and are intended to be acquired in some way by the teacher-learner; and (2) a model in which the teacher participants extract principles of teaching and learning collaboratively through evidence-based activities, such as conducting (and reflecting on) their own educational research. For both models, the goal is teacher learning that leads to more effective practice. The design of the Streamline to Mastery PD program is based on the latter model, with the additional goal of the teacher-learners becoming agents of broader educational change.

One intended outcome of learning in the Streamline to Mastery model is that the teachers increasingly engage in the practice of extracting from their experiences generally applicable principles of teaching and learning that may lead to more effective teaching and broad scale educational change. The process of abstracting general principles from reflection on experiences (including systematic observation and social consensus) is known as the process of *induction* (Bacon, 1878). Induction is a complex and iterative form of reasoning involving moving inferentially from specific instances (concrete experience and observations) of a phenomenon toward generally applicable (typically predictive or mechanistic) rules that govern the behavior of that phenomenon. The inductive process in Streamline to Mastery is analogous to the work of communities of scientists that move inferentially from the observation of specific instances of natural phenomenon to the development of general rules about it. The usefulness of these general rules about our interactions with nature, *scientific principles* and *theories,* resides in the predictive and explanatory power they provide. Likewise, communities of teachers may become both practiced in, and come to identify with, the social practice of using evidence to develop and abstract generally applicable principles of teaching and learning. Just as our empirical work investigates the premise that science learners can author scientific principles and models from evidence (Belleau & Otero, 2012; Ross & Otero, 2011) as they increasingly establish identities associated with science, we investigate here the role of authorship in the development of agency and identities as teacher-leaders.

## Streamline to Mastery

The Streamline to Mastery program began with four science teachers in 2009, grew to nine teachers in 2010, and ten more are to be recruited in the 2014. Requirements to be in the program included teaching in a high needs school district, completion of a master's degree, and a willingness to share aspects of teaching practices through group collaboration. Additionally, teachers are required to conduct research into their own practices, present at least once per year at a national education conference, and take one graduate level college course (of their choice) per year. The research team, all who participated directly in the program, consisted of the NSF project PI, two doctoral students in physics education research (both of whom were formerly high school physics teachers), and one future physics teacher (who was, at the time, an undergraduate Noyce Fellow). Teachers and researchers met every two weeks to share lessons, plan classroom research, and discuss topics of interest to the teachers. Activities included lesson-sharing, in which teachers and researchers each shared a lesson that they deemed to be effective and "inquiry-oriented," designing and executing education research, and preparing to present at national conferences.

Since its inception, the participants of Streamline have struggled with the apparent lack of structure of the program. From the researchers' perspective, it was difficult to balance the need for structure at the outset in order to establish norms for supporting learning with the undergirding philosophy that the teachers know best what their own needs are and are capable of learning to design experiences to meet those needs. Thus, the structures, leadership, and mentoring provided by the principle investigator and two graduate students was deliberately tapered over time. As hypothesized, and as the data below show, the teachers increasingly took control of and responsibility for the design and completion of tasks and for the direction, vision, and mission of the group.

## Extracting Principles of Teaching and Learning

One example of the inductive process that occurred within the Streamline to Mastery community involved the first cohort of four teachers negotiating the meaning of the term "inquiry." In the first weeks of Streamline to Mastery, the teachers agreed that their prior science teacher preparation and professional development had not adequately prepared them to enact effective science instruction. They were all familiar with the word "inquiry," and knew it was somehow important in science learning, yet, just as a great deal of research has shown (Kang, Orgill, & Crippen, 2008; Wallace & Kang, 2004; Windschitl, 2004), all agreed that their understanding of it was inadequate. The term *scientific inquiry*, which is another term for the scientific inductive process described earlier, is a practice that is critical for the authoring of science ideas in the classroom. Though the researchers were quite aware that the term "inquiry" had become so ambiguous and problematic that it would be abandoned in the Next Generation Science Standards (NRC, 2013), we understood that these teachers felt strongly that they needed to understand how to engage their students in *doing* science in the classroom, regardless of what policy document authors decide to call it. Lesson-sharing was one of the structures used by the teachers to explore the notion of *inquiry* as it relates to science learning in the classroom. Each teacher was asked to bring in their best "inquiry lesson" and share it with the group. The group then discussed the ways in which each lesson was consistent with their current ideas of "inquiry." The researchers studied how the teachers' notions of inquiry changed over time throughout this process. Data collected involved periodic reflections, a scenario-based scientific inquiry survey (Kang et al., 2008), and videos of all of the lesson-sharing and other group discussions (for detailed methods, see (Ross, et al, 2011)). These data were coded using both an *a-priori* coding method based upon the elements of classroom scientific inquiry outlined in *Inquiry and the NSES* (National Research Council, 2000) and an open coding method (Strauss, 1987) that allowed for the emergence of codes and trends in the data.

Through a longitudinal analysis of teacher talk, the researchers constructed a time series representation of teachers' collective negotiation of meaning of the term "inquiry" as inferred from the data. The analysis yielded four phases of meaning negotiation. The four phases are shown in Figure 1, along with a representative sample of the transcript excerpts. In the first phase, the teachers used the term "inquiry" in different and ambiguous ways (I), including "hands on," "real-world," "constructivism," and "best practice." In the second phase, the teachers realized and externalized that they did not have a complete understanding of the term (II).

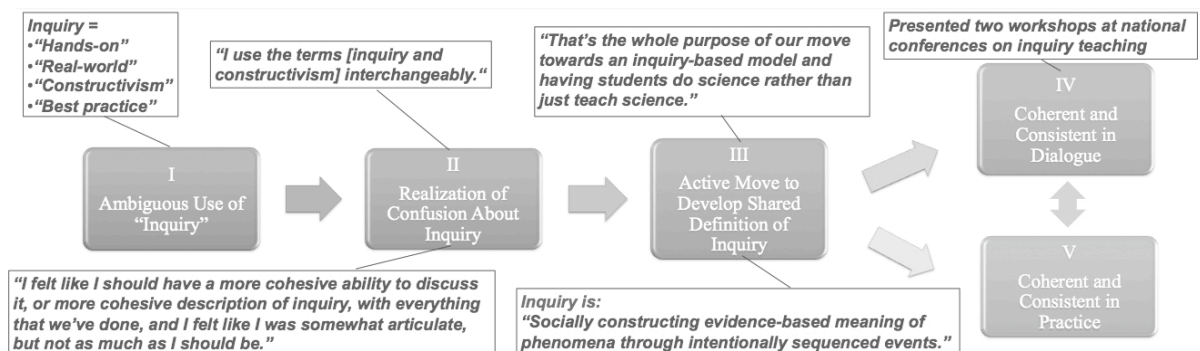

Figure 1. Phases of meaning negotiation of the term "inquiry" among Streamline teachers.

Upon this realization, the teachers decided that they needed to explicitly negotiate better understanding of the meaning and associated practices of inquiry (III). Through concerted effort and communal participation, the group negotiated a shared definition of inquiry as: *"socially constructing evidence-based meaning of phenomena through intentionally sequenced events."* Finally, the teachers used their definition as the foundation for later local and national workshops that they led for other teachers on classroom scientific inquiry (IV). Note that though phase V would be the next logical step in this research, it was not systematically observed or investigated in this study.

The study of the teachers' negotiation of meaning of the term inquiry is an empirical example of the induction of a principle of science teaching and learning. This process was initiated by the teachers and resulted in the extraction of a general principle of teaching and learning. The Streamline teachers moved inferentially from concrete experiences (activity in their own classrooms) and structured observations (lesson-sharing) to a consensus definition of a central topic of interest in science learning. This example of becoming practiced in induction is just one change that suggests the evolution of the Streamline teachers' identities as expert learners who extract general principles from experience and systematic observation.

## Changing Roles, Practices, and Identities

The extraction of principles is one aspect of the learning that was observed among the Streamline to Mastery group. In tandem with the collection of data on the teachers' pursuit of an understanding of "inquiry," group emails and meeting videos were systematically examined for indications of changes in the nature of the teachers' participation as they took on new roles and responsibilities associated with the program. These changes in practice were taken as indicators of increasing agency among the teacher participants. The group emails were analyzed to determine from whom they were sent (teacher or researcher), the date of their origination, what their primary topic was, and if they represented a new line of discussion or were a reply to another email. Transcripts of the meeting videos were coded for the introduction of new agenda items and who introduced them (for detailed methods, see (Van Dusen & Otero, 2012)).

As is shown in Figure 2, the researchers sent the majority of the emails in 2009, but over time the teachers began to send a larger share of the total number of emails. The total number of emails in a given month ranged from eight to fifty four. During period (a) in Figure 2, the majority of the emails were from researchers and focused on scheduling meetings with the occasional email about meeting agenda items. During period (b), the teachers' email volume exceeded those of the researchers. During period, (c), teachers sent an increased percentage of the emails in April. In May, researchers' emails exceeded those of the teachers, largely because the researchers were providing guidance and feedback to teachers as they prepared for their first national conference.

The same set of emails was analyzed to determine whether the teachers or researchers were initiating the *conversation threads*. In this analysis a slightly different pattern emerges. Figure 3 shows the percentage of new email conversation threads by month, again broken down into three time periods. During period (a), researchers initiated nearly all of the new conversations. During period (b), threads initiated by teachers and researchers were evenly balanced with the exception of the March. In March, the teachers were preparing presentations for a regional conference, which required significant communication among the teachers. During this time the researchers primarily acted as resources in answering teacher questions. During period (c), email conversations begun by teachers and researchers were nearly balanced. Figure 2 suggests that teacher participation increased over time, and Figure 3 suggests that the teachers increased their involvement and initiative for leadership.

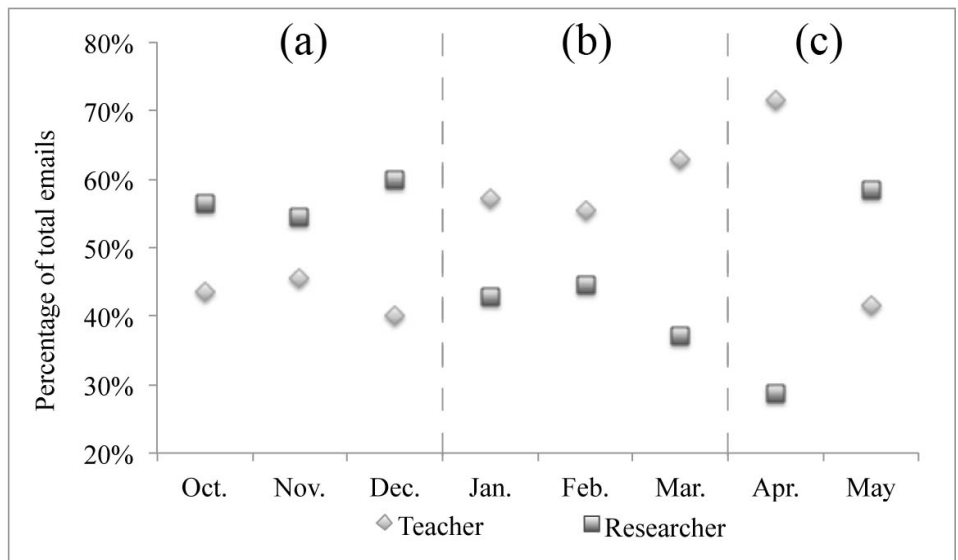

Figure 2. Percentage of total emails sent each month.

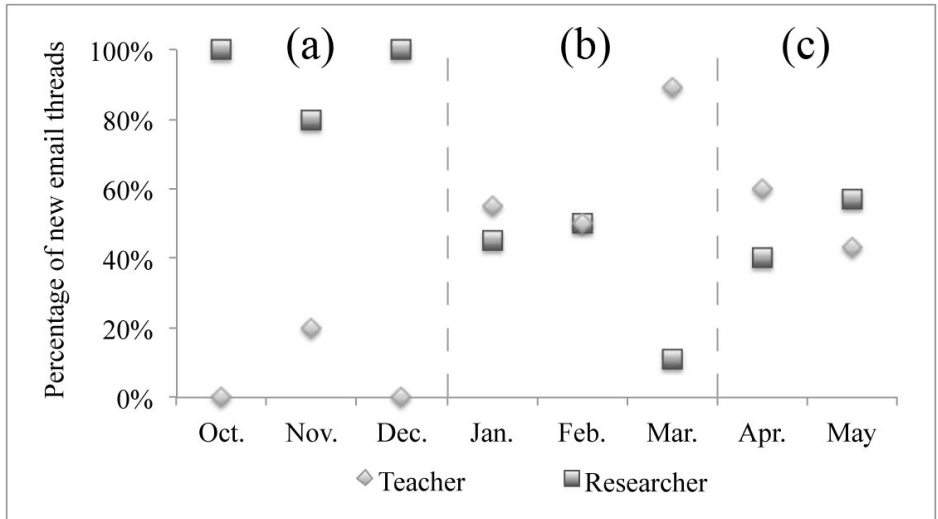

Figure 3. Percent of new email discussion threads.

Videos of group meetings were also analyzed to explore possible changes in who, teachers or researchers, set agendas. We averaged data within each time period (early, middle, and late). As shown in Figure 4(a), researchers provided all of the meeting agenda items during the early time period. The middle time period shows substantial growth in the percentage of agenda items provided by teachers. During the late section (c), the teachers provided the majority of the agenda items. The change that takes place between (a) and (b) came largely from the teachers beginning to take charge of the meetings, ultimately taking on a majority of the agenda setting responsibility.

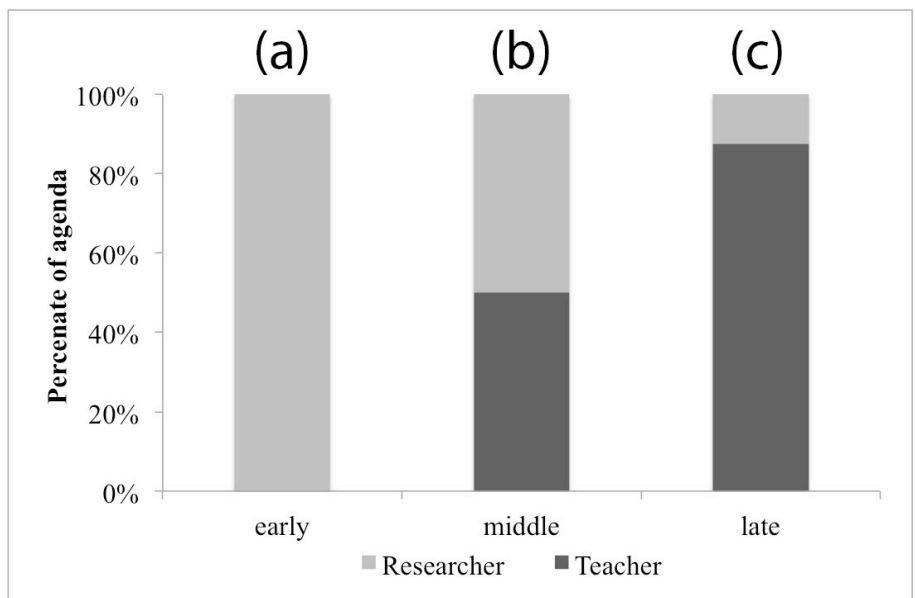

Figure 4. Agenda items from teachers and researchers.

More recently, we examined the type and frequency of the participating teachers' practices of propagating knowledge generated by the Streamline community to other researchers and practitioners. The teachers are responsible for updating a communal spreadsheet with any formal presentations of research findings and teaching practices that they give. This spreadsheet serves as a way for the teachers to keep informed of the work that their peers have done and gives the researchers an easy way to track the community's activities.

As is shown in Figure 5, the instances of sharing knowledge generated within the Streamline community has increased over the four-year period since the group was formally created. Prior to the formation of Streamline, the teachers did not have any role in formally sharing their knowledge of teaching and learning. As a result of being members of the Streamline community, each teachers' formal sharing of knowledge about teaching and learning increased from zero presentations in 2010 to 27 presentations this year (2013), with several more presentations planned before the year ends. As the teachers have taken on stronger teacher-researcher identities, they have also taken up the practice of prioritizing the generation of new knowledge and sharing their learning with others. Not only is the propagation of research findings a practice associated with teacher-leaders and agents of educational change, but the teachers' activity of executing research projects constitute both learning and becoming. Through this research, the teachers generated new knowledge about the topics of study *and* took up identities of those who are competent at learning through rigorous research.

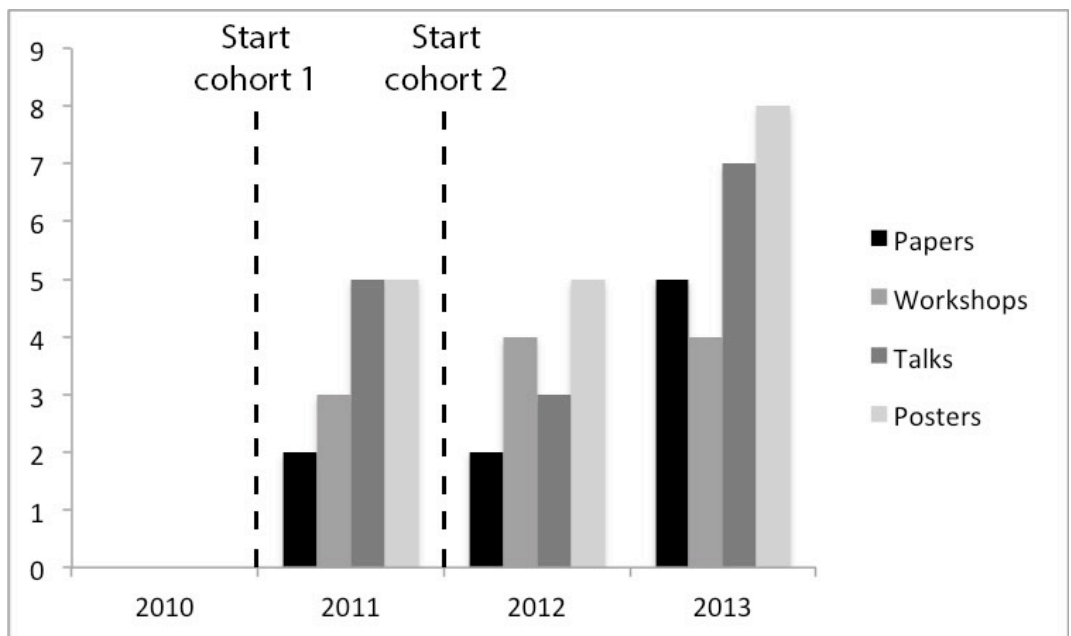

Figure 5. Propagation of Streamline learning to others.

The propagation of knowledge generated through research is tangible evidence of the teacher participants' engagement in the process of induction. This research resulted in the extraction of principles of teaching and learning science. Just as the first cohort of Streamline teachers negotiated meaning of the term "inquiry" in the early stages of the program, all nine participants have engaged in an analogous, and ostensibly more rigorous and systematic, processes of extracting principles of teaching and learning science. Furthermore, many of these findings were recognized as legitimate contributions to the body of education research through peer-review. Moreover, it is the authors' view that the learning and becoming through the disciplinary practices exhibited here may support continued agency in the realms of local and national education policy. The processes of extracting and sharing principles of teaching and learning and of supporting claims about effective policies for education reform with peer-reviewed, evidence-based research has afforded some Streamline to Mastery participants unanticipated levels of agency in national reform efforts. One exemplar of this agency is the participation of a Streamline teacher in the American Chemical Society's (ACS) efforts to launch a chemistry teacher organization comparable to the American Association of Physics Teachers (AAPT). This teacher played a key role in persuading the ACS, through the presentation of her research and experiences with Streamline to Mastery and AAPT, to fund and institutionalize the American Association of Chemistry Teachers, an organization dedicated to the continued development and support of effective chemistry teaching at all levels.

## Conclusions and Implications

In this paper, we have examined the learning of teachers participating in a teacher-driven professional development program. Through a sociocultural perspective of identity development, these data demonstrate that Streamline teachers became practiced in various ways that suggest the development of identities as evidence-based advocates of educational change. By extracting their own principles of teaching and learning through lesson-sharing, reflection, and research, the teachers participated in communities that identify with the practices of using evidence to inform and advocate for science education reform. The abstraction of principles of teaching and learning and the taking on of identities of evidence-based education reformers is indeed an example of *learning and becoming in practice.*

It is no coincidence that the model of teacher learning that we espouse is modeled after the central practice of science, *induction.* Just as the core practice of science is to infer from evidence and abstract through consensus the principles, theories, and models that predict, explain, and fundamentally change the ways that humans understand and interact with the natural world, we see generality in this practice. This process of generating new knowledge is applicable to teachers' generation of knowledge as well as formal and informal K-

16 science learning. In our interpretations and discussions of data on learner-driven models of PD and science learning, we must point out what we see as the critical role of the fuller participants (Lave & Wenger, 1991) in these activities. In our view, the development of principles in science, science learning, and teaching and learning is guided heavily by the expectations, norms, and pedagogical expertise of the more knowledgeable others of a community. Just as effective science teachers must be able to engage students in the practice of scientific induction, administrators of PD must strike some productive balance between providing structure and guidance to the learners and providing space and opportunities for the expressions of agency and the identity development valued by the designers and participants in these contexts.

The body of research we are generating suggests that authority-based models of learning science and of learning the practices of effective science teaching may never realize the agency and identity development that is associated with learning in which the making of inferences from evidence and the abstraction through consensus and subsequent reification of these inferences are the domain of the learners. We must continue to explore the premise that for meaningful learning to occur, the very development of abstract, general principles must be performed by the learners, as opposed to delivered to them by some authority, whether it be text, teacher, or some digital medium. Of course, these broad claims warrant further study, but we assert that the learning demonstrated by the data presented about Streamline to Mastery supports the assertion that learner-driven models of induction can be associated with marked changes in agency in valued practices. As the Next Generation Science Standards (NRC, 2013) impose a new set of demands on current and future science teachers, we can be reasonably sure that the dominant, top-down models of teacher (and student) learning will be inadequate to the task of supporting teachers in enacting these standards. Learning through a general model of learner-driven, evidence-based induction has the potential to clarify the central goals of science education, science teacher education, and to more accurately represent science in our classrooms.

## References

Bacon, F. (1878). *Novum Organum*. Oxford: Clarendon Press.
Bahktin, M. (1993). *Toward a Philosophy of the Act*. Austin, TX: University of Texas Press.
Barton, A., & Tan, E. (2010). We be burnin'! Agency, identity, and science learning. *The Journal of the Learning Sciences*, *19*(2), 187–229.
Belleau, S., Otero, V. (2012). Critical Classroom Structures for Empowering Students to Participate in Science. In S. Rabelo, C. Singh, & P. Engelhardt (Eds.) *2012 Physics Education Research Conference Proceedings.* Melville, NY: AIP Press.
Borko, H. (2004). Professional Development and Teacher Learning: Mapping the Terrain. *Educational Researcher*, *33*(8), 3–15. doi:10.3102/0013189X033008003
Desimone, L. M. (2009). Improving Impact Studies of Teachers' Professional Development: Toward Better Conceptualizations and Measures. *Educational Researcher*, *38*(3), 181–199. doi:10.3102/0013189X08331140
Garet, M. S., Porter, a. C., Desimone, L., Birman, B. F., & Yoon, K. S. (2001). What Makes Professional Development Effective? Results From a National Sample of Teachers. *American Educational Research Journal*, *38*(4), 915–945. doi:10.3102/00028312038004915
Grossman, P., Wineburg, S., & Woolworth, S. (2001). Toward a Theory of Teacher Community. *Teachers College Record*, 942–1012.
Holland, D., Lachicotte, W., Skinner, D., Cain, C. (1998). *Identity and Agency in Cultural Worlds*. Cambridge, MA: Harvard University Press.
Kang, N.-H., Orgill, M., & Crippen, K. J. (2008). Understanding Teachers' Conceptions of Classroom Inquiry With a Teaching Scenario Survey Instrument. *Journal of Science Teacher Education*, *19*(4), 337–354. doi:10.1007/s10972-008-9097-4
Lave, J., Wenger, E. (1991). *Situated Learning: Legitimate Peripheral Participation*. Cambridge: Cambridge University Press.
Nasir, N. S., & Hand, V. M. (2006). Exploring Sociocultural Perspectives on Race, Culture, and Learning. *Review of Educational Research*, *76*(4), 449–475. doi:10.3102/00346543076004449
Nasir, Na'ilah Suad. (2002). Identity, Goals, and Learning: Mathematics in Cultural Practice. *Mathematical Thinking and Learning*, *4*(2-3), 213–247. doi:10.1207/S15327833MTL04023_6
National Research Council. (2013). *Next Generation Science Standards: For States, By States*. Washington DC: The National Academies Press.
National Research Council. (2000). *Inquiry and the National Science Education Standards: A Guide for Teaching and Learning*. *Science Education*. Washington D.C.: National Academies Press.

Ross, M. & Otero, V. (2012). Challenging Traditional Assumptions of High School Science through the PET Curriculum, in S. Rabelo, C. Singh, & P. Engelhardt (Eds.) *2012 Physics Education Research Conference Proceedings*. Melville, NY: AIP Press.

Ross, M., Van Dusen, B., Sherman, S., & Otero, V. (2011). Teacher-driven professional development and the pursuit of a sophisticated understanding of inquiry. In Rebello, N. S., Engelhardt, P. V., & Singh, C. (Ed.), *Physics Education Research Conference* (pp. 327–330). doi:10.1063/1.3680061

Strauss, A. L. (1987). *Qualitative analysis for social scientists*. Cambridge Univ Press

Van Dusen, B., Otero, V. (2012). Influencing Student Relationships With Physics Through Culturally Relevant Tools. In P. E. S. Rabello, C. Singh (Ed.), *Physics Education Research Conference Proceedings*. Melville, NY: AIP Press. Retrieved from http://arxiv.org/abs/1209.1365

Wallace, C., & Kang, N. (2004). An investigation of experienced secondary science teachers' beliefs about inquiry: An examination of competing belief sets. *Journal of Research in Science Teaching*, *41*(9), 936–960. doi:10.1002/tea.20032

Windschitl, M. (2004). Folk theories of inquiry: How preservice teachers reproduce the discourse and practices of an atheoretical scientific method. *Journal of Research in Science Teaching*, *41*(5), 481–512. doi:10.1002/tea.20010